\documentclass[prd,aps,nofootinbib,superscriptaddress,floatfix,10 pt]{revtex4}%
\usepackage{amssymb}
\usepackage{amsmath,graphicx,color,epstopdf,graphics}
\usepackage{subfigure}
\usepackage{amsmath}
\usepackage{amsfonts}
\usepackage{graphicx}%
\begin{document}
\preprint{ }
\title{Coherent Destruction of Tunneling in a\ Two Electron Double Quantum Dot:
Interplay of Coulomb interaction, spin-orbit interaction and AC field }
\author{K. Rahim}
\affiliation{Department of Physics, Quaid-i-Azam University, Islamabad, Pakistan}
\author{U. Hasan}
\affiliation{Department of Physics,Quaid-i-Azam University and National Center of Physics,
Islamabad, Pakistan}
\author{K. Sabeeh}
\affiliation{Department of Physics, Quaid-i-Azam University, Islamabad, Pakistan}
\author{}
\affiliation{}
\author{}
\affiliation{}

\begin{abstract}
We study a double quantum dot system with two interacting electrons in the
presence of a time-dependent periodic (AC) electric field and spin-orbit
interaction. We focus on the phenomenon of Coherent Destruction of Tunneling
(CDT) for an initially localized state. Because of the periodicity introduced
by the AC field we use Floquet theory to find quasi energies with their
crossing and anti-crossing points corresponding to CDT . We observe that the
AC field rescales the spin orbit and hopping amplitudes in terms of Bessel
functions. The zeros of the Bessel functions are of the form of a ratio of AC
field strength to its frequency and quasi energies at these points form
anti-crossings in our case. We first prepare the system in a triplet state and
study the evolution of its probability in the presence of spin-orbit
interaction alone. We observe an oscillatory behavior which indicates spin
flip transitions. However when the AC field is turned on the probability
oscillations are highly suppressed at anti-crossing points leading to
localization of initial state and the state retains its memory of spin even in
the presence of spin flip interaction.

\end{abstract}
\startpage{01}
\endpage{02}
\maketitle

\section{Introduction}

Semiconductor quantum dots are small conductive regions in a semicoductor with
tunable number of electrons. Their shape and size can be varied by varing the
gate voltage. Quantum dots are potential candidates for building blocks of a
quantum computer \cite{1,2}. For this to work, it is necessary to couple the
dots coherently and keep this coherence for longer times. In this regard, spin
dephasing time is of special interest. Studies of spin dephasing time have
been carried out in the presence of a magnetic field in \cite{3,4} on single
quantum dots(QDs). Double Quantum Dots (DQDs), where tunneling between the
dots is an important additional parameter, introduce new energy scales which
open avenues of rich physics with possible applications in spintronics. In
DQDs, the spins can be manipulated in many ways such as by electric fields in
the presence of a constant magnetic field, taking advantage of both spin-orbit
and nuclear hyperfine interactions \cite{5,6,7,8,9}. In these systems, spin
manipulation in the presence of AC electric field has also been investigated
by considering the periodic motion of the electron leading to electron dipole
spin resonance(ESDR) \cite{10}. Prior to our work, spin manipulation of single
electron in a QD has been studied in the presence of radiation in THZ
range\cite{11}; the authors have investigated the electronic Density of States
(DOS) in the presence of both spin-orbit interaction and electromagnetic
radiation. Other studies consider the effect of an external oscillating
electric field on Rabi spin oscillation of a single electron in a DQD
\cite{12}. This work shows that Rabi spin oscillation can be slowed at high
electric field. They have also shown a decrease of displacement of electron
which results in slow driving of spin with increase in electric field. This
shows strong dependence of efficiency and time scale of spin manipulation on
the external field.

In this work, we investigate spin flip processes due to spin-orbit interaction
in a DQD system with two interacting electrons in the presence of external AC
electric field. Spin-orbit interaction is the main source of spin flip for the
three- and two-dimensional electron states in GaAs-type crystals \cite{13,14}.
The unit cell in this material has no inversion center which gives rise to an
effective spin-orbit coupling in the electron spectrum. We show that Coherent
Destruction of Tunneling(CDT) occurs in our system. The observation of CDT
means suppression of spin flip processes which can be attributed to the AC
field localizing the initial state. In this regard our results present a
different and more general view of spin dynamics compared to \cite{12} where
only a single electron is considered and spin oscillation suppression is observed.

Because of the periodicity introduced by the externally applied field Floquet
formalism can be used to investigate the evolution of the system. The
quasi-energies are calculated by finding the one period propagator and the
dynamics of the system are determined for time scale much larger than the
driving period. On this time scale the behavior of the system is completely
described by the quasi-energies. In the presence of electric field alone, the
dynamics of the single electron in a DQD have been rigorously studied with the
successful use of this technique \cite{15}. The effect of crossing and
anti-crossing of quasi-energies on the dynamics of electrons was studied in
\cite{16} and \cite{17,18}. However, any such scheme based on Hubbard model
separates the singlet and triplet subspaces and consequently the dynamics does
not show spin non-conserving transitions. The presence of spin-orbit
interaction allows for mixing of singlet and triplet states which are
completely decoupled otherwise. We choose spin-orbit field direction such that
z axis in spin space lies parallel to spin-orbit field, $\boldsymbol{\Omega}$;
in which case two of the triplet states are decoupled from the remaining spin
states \cite{19,20}. We show that the effect of this field is to rescale
tunneling (hopping) and spin-orbit interaction parameters in terms of Bessel
functions. This rescaling depends on initial values of these parameters and
ratio of electric field strength to its frequency. This leads to CDT of the
initial state when the ratio of field strength and frequency is tuned to zeros
of the Bessel function.

We show that the initial state prepared as a spin triplet can transition into
singlet subspace due to spin-orbit coupling. Our main focus is to see how this
transition can be suppressed by the application of an external electric field.
This we relate to the behavior of the initial state at the points near the
anti crossings.

This paper is organized as follows. In Sec. II, we develop the model for two
interacting electrons confined in a DQD and define a suitable basis for the
interacting electrons. In Sec. III, we introduce spin-orbit interaction in the
second quantized formalism and define an appropriate geometery for our model.
In the next section we give a brief introduction to Floquet theory. Using this
formalism we then define Floquet states and quasi-energies in Sec. V. We
investigate crossings and anti crossings of quasi-energies and study CDT
corresponding to these points. Parameters that tune CDT leading to
localization of electrons in our system are discussed. For this purpose, we
prepare the system in an initial state and study its time evolution. Finally,
we present our results in Sec. VI.

\section{Model}

We consider two electrons confined in a $DQD$ system, Fig.1. The dots lie in
the plane $(\xi,$z) with tunneling that can occur in the $\xi$ direction. The
spin-orbit field $\boldsymbol{\Omega}$ points along the z axis which also
defines the spin quantization axis. The dots are detuned by externally applied
voltage$.$ Fig.\ref{fig:2} shows the effect of detuning on the quantum dot
levels. We assume that the electrons are confined near the minima of a double
well potential $V_{QD}$ created by electrical gating of the system. Since
there are two electrons in the system, the most suitable basis is $(n,m)$
where $n$ and $m$ denote the number of electrons in the left and right dot
respectively. In the second quantized notation, including spin, the states can
be written as
\begin{align}
|(2,0)S\rangle &  =c_{L\uparrow}^{\dagger}c_{L\downarrow}^{\dagger}%
|0\rangle\label{1}\\
|(0,2)S\rangle &  =c_{R\uparrow}^{\dagger}c_{R\downarrow}^{\dagger}%
|0\rangle\label{2}\\
|(1,1)S\rangle &  =\frac{1}{\sqrt{2}}(c_{L\uparrow}^{\dagger}c_{R\downarrow
}^{\dagger}-c_{L\downarrow}^{\dagger}c_{R\uparrow}^{\dagger})|0\rangle
\label{3}\\
|T_{+}\rangle &  =c_{L\uparrow}^{\dagger}c_{R\uparrow}^{\dagger}%
|0\rangle\label{4}\\
|T_{0}\rangle &  =\frac{1}{\sqrt{2}}(c_{L\uparrow}^{\dagger}c_{R\downarrow
}^{\dagger}+c_{L\downarrow}^{\dagger}(c_{R\uparrow}^{\dagger})|0\rangle
\label{5}\\
|T_{-}\rangle &  =c_{L\downarrow}^{\dagger}c_{R\downarrow}^{\dagger}%
|0\rangle\label{6}%
\end{align}
where $S$ and $T$ donate the singlet and triplet spin states. $c_{i}^{\dagger
}(c_{i})$ are the usual creation (annihilation) operators. The orbital parts
of the wavefunction for $|(1,1)S\rangle$ and $|T_{0,\pm}\rangle$ are given by
\begin{equation}
\Psi_{\pm}^{s}(r_{1},{r_{2}})=\frac{1}{\sqrt{2}}{[\Phi_{L}(r_{2})\Phi
_{R}(r_{2})\pm\Phi_{R}(r_{1})\Phi_{L}(r_{2})]} \label{7}%
\end{equation}
The orbital parts of the other two singlet states $|(0,2)S\rangle$ and
$|(2,0)S\rangle$, which represent double occupation on right and left dot
respectively, are given by%
\begin{equation}
\Psi_{L,R}^{d}(r_{1},{r_{2}})=\Phi_{L,R}(r_{1})\Phi_{L,R}(r_{2}) \label{8}%
\end{equation}
The orbital functions $\Psi_{+}^{s}(r_{1},{r_{2}})$ and $\Psi_{L,R}^{d}$ are
symmetric under exchange of particles while $\Psi_{-}^{s}(r_{1},{r_{2}})$ is
antisymmetric under exchange. $\Phi_{L}$ and $\Phi_{R}$ are Wannier orbitals
centered on the left and right dot respectively \cite{21}. In our model the
energies of the interacting electrons are detuned by an amount $2\varepsilon$.
This detuning potential $V_{bias}$ can be established by the electrostatic
gates which also create $V_{QD}$. The detuning $\varepsilon$ represents an
energy difference for an electron occupying the left or the right dot. For the
symmetric case the electrostatic gates voltages are adjusted such that an
electron would have same energy in either of the two wells, see
Fig.\ref{fig:2}. The electrons in the DQD are coupled through Coulomb
interaction $U$ which is the cost of double occupation in states
$|(2,0)S\rangle$ and $|(0,2)S\rangle$. Hopping between the dots is given by
$W$ and repulsion of singlet(triplet) state with single electron in each dot
is given by $V_{+}(V_{-})$. In addition, we subject the DQD system to an AC
electric field in the presence of spin orbit interaction. The oscillating
electric field can be established by the electrical gates to which an AC
signal is supplied from external source. The AC field, confinement potential
$V_{QD}$, Coulomb interaction, the detuning $\varepsilon$ and hopping
constitute the spin independent part of the Hamiltonian:
\begin{equation}
H_{0}=\left(  -1\right)  ^{i}%
%TCIMACRO{\dsum \limits_{i,\sigma}}%
%BeginExpansion
{\displaystyle\sum\limits_{i,\sigma}}
%EndExpansion
(\varepsilon+E(t))c_{i,\sigma}^{\dagger}c_{i,\sigma}+W%
%TCIMACRO{\dsum \limits_{\sigma}}%
%BeginExpansion
{\displaystyle\sum\limits_{\sigma}}
%EndExpansion
(c_{L,\sigma}^{\dagger}c_{R,\sigma}+c_{R,\sigma}^{\dagger}c_{L,\sigma})+U%
%TCIMACRO{\dsum \limits_{i,\sigma}}%
%BeginExpansion
{\displaystyle\sum\limits_{i,\sigma}}
%EndExpansion
n_{i,\uparrow}n_{i,\downarrow} \label{9}%
\end{equation}
where we have introduced the spin index $\sigma$. The summation on $i$ runs
over left and right dots, i.e $1,2$ respectively. $E(t)=Vcos{(\omega t)}$
represents the time dependent electric field with $V$ and $\omega$ represents
the strength and frequency of the oscillating field respectively.

\section{The Spin-orbit interaction}

In the basis defined in Eqns.(\ref{1})$-$(\ref{6}), the Hamiltonian,
Eqn.(\ref{9}), can be expressed in the singlet and triplet blocks:
\begin{equation}
H_{0}=%
\begin{pmatrix}
H_{SS} & 0\\
0 & H_{TT}%
\end{pmatrix}
\end{equation}
where the singlet Hamiltonian $H_{SS}$ in the basis $|(2,0)S\rangle$,
$|(0,2)S\rangle$ and $|(1,1)S\rangle$ is%

\begin{equation}
H_{SS}=%
\begin{pmatrix}
U-\varepsilon-Vcos{\omega t} & \sqrt{2}W & 0\\
0 & U+\varepsilon+Vcos{\omega t} & \sqrt{2}W\\
\sqrt{2}W & \sqrt{2}W & V_{+}%
\end{pmatrix}
\end{equation}
and $H_{TT}$ represents the diagonal triplet subspace Hamiltonian.

The transitions between singlet and triplet states can be mediated by the
spin-orbit interaction which can be expressed in second quantized
form\cite{19}. The spin-orbit interaction within the space of low energy
single electron orbitals in DQD is expressed in terms of the spin-orbit field
$\boldsymbol{\Omega}$,%
\begin{equation}
H_{SO}=\boldsymbol{i}\text{ }\boldsymbol{\Omega}\cdot%
%TCIMACRO{\tsum \limits_{\alpha,\beta=\uparrow\downarrow}}%
%BeginExpansion
{\textstyle\sum\limits_{\alpha,\beta=\uparrow\downarrow}}
%EndExpansion
(c_{\alpha,L}^{\dagger}\sigma^{\alpha\beta}c_{\beta,R}-H.C)
\end{equation}
where $\boldsymbol{\Omega}$ is the spin orbit field and is given by
\begin{equation}
\boldsymbol{i}\text{ }\boldsymbol{\Omega}=\langle{\Phi_{L}}\lvert{P_{\xi}%
}\rvert{\Phi_{R}}\rangle{a_{\boldsymbol{\Omega}}}%
\end{equation}
It depends on the orientation of the dots with respect to crystallographic
axes through the vector $a_{\boldsymbol{\Omega}}$\cite{22,23}. $\sigma$
represents the Pauli spin matrix. For a 2DEG in the $(001)$ plane,
$a_{\boldsymbol{\Omega}}$ is given by%
\begin{equation}
a_{\boldsymbol{\Omega}}=(\beta-\alpha)cos{\theta}\boldsymbol{e_{[\overline
{1}10]}}+(\beta+\alpha)sin{\theta}\boldsymbol{e_{[110]}}%
\end{equation}
where the angle between $\boldsymbol{e_{\xi}}$ direction and the $[110]$
crystallographic axis is denoted by $\theta$. The matrix element of $P_{\xi}$,
the momentum component along the $\xi$ direction that connects the two dots,
is taken between the corresponding Wannier orbitals, and it depends on the
envelope wave function and the double dot binding potential \cite{9}. The
spin-orbit interaction causes transition between singlet and triplet states
taking the single occupation of both dots to double occupation of either of
the two dots. The explicit expression for $\boldsymbol{\Omega}$ is \cite{21},
\begin{equation}
\boldsymbol{\Omega}=\dfrac{4W}{3}\dfrac{l}{\Lambda_{SO}}\dfrac
{a_{\boldsymbol{\Omega}}}{|{a_{\boldsymbol{\Omega}}}|} \label{15}%
\end{equation}
where $l$ is inter-dot distance, $\Lambda_{SO}$ is the spin-orbit length.
Inter-dot distance and hopping amplitude are DQD geometry dependent parameters
where as the spin-orbit length $\Lambda_{SO}$ is determined by the material
properties and by the orientation of the DQD with respect to the
crystallographic axis. Introducing spin-orbit interaction in the Hamiltonian
Eqn.(\ref{9}), we obtain the complete Hamiltonian for our system
\begin{align}
H(t)  &  =H_{0}+H_{SO}\label{16}\\
&  =\left(  -1\right)  ^{i}%
%TCIMACRO{\dsum \limits_{i,\sigma}}%
%BeginExpansion
{\displaystyle\sum\limits_{i,\sigma}}
%EndExpansion
(\varepsilon+E(t))c_{i,\sigma}^{\dagger}c_{i,\sigma}+W%
%TCIMACRO{\dsum \limits_{\sigma}}%
%BeginExpansion
{\displaystyle\sum\limits_{\sigma}}
%EndExpansion
(c_{L,\sigma}^{\dagger}c_{R,\sigma}+c_{R,\sigma}^{\dagger}c_{L,\sigma})+U%
%TCIMACRO{\dsum \limits_{i,\sigma}}%
%BeginExpansion
{\displaystyle\sum\limits_{i,\sigma}}
%EndExpansion
n_{i,\uparrow}n_{i,\downarrow}\nonumber\\
&  +\boldsymbol{i}\text{ }\boldsymbol{\Omega}\cdot%
%TCIMACRO{\dsum \limits_{\alpha,\beta=\uparrow\downarrow}}%
%BeginExpansion
{\displaystyle\sum\limits_{\alpha,\beta=\uparrow\downarrow}}
%EndExpansion
(c_{\alpha,L}^{\dagger}\sigma^{\alpha\beta}c_{\beta,R}-h.c).\nonumber
\end{align}
With the z-axis taken along $\boldsymbol{\Omega}$ the Hamiltonian given in
Eqn.(\ref{16}) becomes
\begin{equation}
H(t)=%
\begin{pmatrix}
U-\varepsilon-Vcos{\omega t} & 0 & \sqrt{2}W & 0 & -\iota\sqrt{2}\Omega & 0\\
0 & U+\varepsilon+Vcos{\omega t} & \sqrt{2}W & 0 & -\iota\sqrt{2}\Omega & 0\\
\sqrt{2}W & \sqrt{2}W & V_{+} & 0 & 0 & 0\\
0 & 0 & 0 & \varepsilon_{T{+}} & 0 & 0\\
\iota\sqrt{2}\Omega & \iota\sqrt{2}\Omega & 0 & 0 & V_{-} & 0\\
0 & 0 & 0 & 0 & 0 & \varepsilon_{T_{-}}%
\end{pmatrix}
\label{17}%
\end{equation}
where $\varepsilon_{T{+}}$ and $\varepsilon_{T{-}}$ refer to energies of
triplet states $|T_{+}\rangle$ and $|T_{-}\rangle$ which are degenerate in
energy and do not couple to singlet subspace and $|T_{0}\rangle$ state, due to
our choice of the geometry of the dots and spin-orbit field
$\boldsymbol{\Omega}$ axis. In case of a choice of other directions of
$\boldsymbol{\Omega,}$ the singlet subspace would couple with $|T_{+}\rangle$,
$|T_{-}\rangle$ also. In the block diagonal form we have
\begin{equation}%
\begin{pmatrix}
H_{1}{(t)} & 0\\
0 & H_{2}%
\end{pmatrix}
\label{18}%
\end{equation}
The matrix $H_{1}{(t)}$ is
\begin{equation}%
\begin{pmatrix}
U-\varepsilon-Vcos{\omega t} & 0 & \sqrt{2}W & -i\sqrt{2}\Omega\\
0 & U+\varepsilon+Vcos{\omega t} & \sqrt{2}W & -i\sqrt{2}\Omega\\
\sqrt{2}W & \sqrt{2}W & V_{+} & 0\\
i\sqrt{2}\Omega & i\sqrt{2}\Omega & 0 & V_{-}%
\end{pmatrix}
\label{19}%
\end{equation}
$H_{1}{(t)}$ consists of the singlet states $|(2,0)S\rangle$,$|(0,2)S\rangle
$,$|(1,1)S\rangle$ and $|{T_{0}}\rangle$ whereas $H_{2}$ is a $2\times2$
matrix in the basis of triplet states $|{T_{+}}\rangle$ and $|{T_{-}}\rangle$.

\section{Floquet Formalism}

The Hamiltonian in Eqn.(\ref{19}) is periodic in time: $H_{1}{(t+T)}=H_{1}%
{(t)}$ with $T$ being the period of the driving field. Floquet formalism can
be used to express solutions of the time-dependent Schrodinger equation as
$\Psi_{\alpha}{(x,t)}=\exp({-i\varepsilon_{\alpha}t)}\phi_{\alpha}{(x,t)}$,
where $\phi_{\alpha}{(x,t)}$ is a function with the same periodicity as
$H_{1}{(t)}$ and is called a Floquet state and $\varepsilon_{\alpha}$ is
termed the quasi-energy. The Floquet states and their quasi-energies can be
obtained from the eigenvalue equation:%

\begin{equation}
H_{F}\phi_{\alpha}{(x,t)}=\varepsilon_{\alpha}\phi_{\alpha}{(x,t)} \label{20}%
\end{equation}
where $H_{F}=(H_{I}{(t)}-i\tfrac{\partial}{\partial{t}})$ is called Floquet
Hamiltonian. It is evident that the Floquet states are periodic in time with
the same period as the Hamiltonian. The periodicity of $\phi_{\alpha}{(x,t)}$
allows a series of Floquet eigenvalues, $\varepsilon_{\alpha}+n\hbar\omega,$
with corresponding eigenfunctions $\phi_{n\alpha}{(x,t)}=e^{(in\omega t)}%
\phi_{\alpha}{(x,t)}$, where $n$ is an integer. However, the physical state
$\Psi_{\alpha}{(x,t)}$ is unchanged; the Floquet eigenvalues associated with
distinct physical states are defined only modulo $\hbar\omega$. The
quasi-energies can be obtained by numerical diagonalization of the unitary
evolution operator $U{(T,0)}$ for one period of the field. The eigenvalues of
$U{(T,0)}$ are then related to quasi-energies as $\lambda_{\alpha}%
=\exp[-i\varepsilon_{\alpha}T]$. The time periodicity of the Floquet states
allows us to study the dynamics of the system for time scales larger than the
period of the driving field effectively in terms of quasi-energies. As the
quasi-energies approach degeneracy, the dynamics of the system are frozen,
producing CDT \cite{18}.

\section{Calculation and Results}

To find the approximate solution of Eqn.(\ref{19}) we follow the
perturbation scheme of \cite{24}. We divide the Hamiltonian Eqn.(\ref{18})
into two parts: $H_{I}{(t)}$ which contains the electric field, detuning and
Coulomb terms and $H_{t}$ which contains the tunneling component and
spin-orbit interaction. We proceed by first finding the eigensystem of the
operator $(H_{I}{(t)}-i\tfrac{\partial}{\partial{t}})$ and treating $H_{t}$ as
a perturbation. The advantage of this approach is that the Floquet states are
eigenstates of $(H_{I}{(t)}-i\tfrac{\partial}{\partial{t}})$ and satisfy the
following equation: ${[H_{I}{(t)}-i\frac{\partial}{\partial{t}}}]\phi_{\alpha
}{(x,t)}=\varepsilon_{\alpha}\phi_{\alpha}{(x,t)}$. Hence, the corrections can
be easily evaluated by working in an extended Hilbert space of $T$-periodic
functions \cite{24} by using standard Rayleigh-Schrodinger perturbation
theory. In our basis $H_{I}{(t)}$ is diagonal and the resulting orthonormal
set of eigenvectors for $(H_{I}{(t)}-i\frac{\partial}{\partial{t}})$ are given
by
\begin{align}
|u_{1}{(t)}\rangle &  =\Bigg(exp\bigg[-i\big(U-\varepsilon-\varepsilon
_{1}\big)t-i{\tfrac{V}{\omega}}sin{(\omega t)}\bigg],0,0,0\Bigg)\label{21}\\
|u_{2}{(t)}\rangle &  =\Bigg(0,exp\bigg[-i\big(U+\varepsilon-\varepsilon
_{2}\big)t-i{\tfrac{V}{\omega}}sin{(\omega t)}\bigg],0,0\Bigg)\label{22}\\
|u_{3}{(t)}\rangle &  =\Bigg(0,0,exp\Big(i\varepsilon_{3}%
t\Big),0\Bigg)\label{23}\\
|u_{4}{(t)}\rangle &  =\Bigg(0,0,0,exp\Big(i\varepsilon_{4}%
t\Big)\Bigg) \label{23b}%
\end{align}
Using the $T$-periodicity, the eigenvalues $\varepsilon_{3,4}$ are zero and
$\varepsilon_{1}=U+\varepsilon$ and $\varepsilon_{2}=U-\varepsilon$. These
eigenvalues represent zeroth order approximation to the quasi-energies. An
interesting feature of these values is that for the choices $U=\varepsilon$
and $2U=n\omega,$ all four energies are degenerate. However, this degeneracy
is lifted by $H_{t}$ which is treated as a perturbation. The first order
correction to the quasi-energies can be calculated in the extended Hilbert
space of time periodic functions by defining an appropriate scalar product
$P_{ij}=\left\langle \left\langle u_{i}\left\vert H_{t}\right\vert
u_{i}\right\rangle \right\rangle ,$ where $\left\langle \left\langle
...\right\rangle \right\rangle $ denotes the inner product in the extended
Hilbert space. It is straight forward to calculate matrix elements of $P$ and
subsequently its eigen values can be found. The matrix form of $P$ is found to be:%

\begin{equation}%
\begin{pmatrix}
0 & 0 & \sqrt{2}W(-1)^{n}J_{n}{(\beta)} & -i\sqrt{2}\Omega(-1)^{n}J_{n}%
{(\beta)}\\
0 & 0 & \sqrt{2}WJ_{n}{(\beta)} & -i\sqrt{2}\Omega J_{n}{(\beta)}\\
\sqrt{2}W(-1)^{n}J_{n}{(\beta)} & \sqrt{2}WJ_{n}{(\beta)} & V_{+} & 0\\
i\sqrt{2}\Omega(-1)^{n}J_{n}{(\beta)} & i\sqrt{2}\Omega J_{n}{(\beta)} & 0 &
V_{-}%
\end{pmatrix}
\label{24}%
\end{equation}
where $\beta=\tfrac{V}{\omega}$. Thus we note that the electric field has
redefined the tunneling and spin-orbit interaction in terms of the Bessel
functions $J_{n}{(\beta)}$. The eigen values of the $P$ give first order
approximationto quasi-energies. Exact or near degeneracy of the quasi-energies
results in the suppression of tunneling of the associated Floquet states,
leading to localization of the initial state. This occurs whenever the ratio
$\tfrac{V}{\omega}$ is equal to the root of the Bessel function.
Fig.\ref{fig:3(b)} shows the locations of the quasi-energies as a function of
$\tfrac{V}{\omega}$ from this perturbative treatment with $n=2$.

Now we study the interplay between spin dynamics and the AC field we consider
the time evolution of an initially localized state. We choose the system to be
initially in the triplet state $|{T_{0}}\rangle$. The time evolution of an
initial state is described by the Schrodinger equation. The wave function can
be written as%

\begin{equation}
|{\Psi{(t)}}\rangle=C_{1}{(t)}|(2,0)S\rangle+C_{2}{(t)}|(0,2)S\rangle
+C_{3}{(t)}|(1,1)S\rangle+C_{4}{(t)}|T_{0}\rangle
\end{equation}
Focusing on $H_{1}{(t),}$ the evolution of the system is given by
\begin{align}
\dot{iC_{1}}{(t)}  &  =(U-\varepsilon-Vcos{(\omega t)})C_{1}{(t)}+\sqrt
{2}WC_{3}{(t)}-i\sqrt{2}\Omega C_{4}{(t)}\label{25}\\
i\dot{C_{2}}{(t)}  &  ={(U+\varepsilon+Vcos{(\omega t)})C_{2}{(t)}+\sqrt
{2}WC_{3}{(t)}-i\sqrt{2}\Omega C_{4}{(t)}}\label{26}\\
\dot{iC_{3}}{(t)}  &  =\sqrt{2}WC_{1}{(t)}+\sqrt{2}WC_{2}{(t)}+V_{+}C_{3}%
{(t)}\label{27}\\
i\dot{C_{4}}{(t)}  &  =i\sqrt{2}\Omega C_{1}{(t)}+i\sqrt{2}\Omega C_{2}%
{(t)}+V_{-}C_{4}{(t)} \label{28}%
\end{align}
This system of equations can be solved numerically with the initial conditions
$\big(C_{1}{(0)},C_{2}{(0)},C_{3}{(0)},C_{4}{(0)}=0,0,0,1\big)$. We will
define the minimum value of $|{C_{4}{(t)}}|^{2}$ in 20 driving periods as
$P_{(min)}^{4}$. Hence the case $P_{(min)}^{4}=1$ means that the system can
maintain its initial state. In this case, tunneling between different states
is completely suppressed. The case $P_{(min)}^{4}=0$ means that the initial
state cannot be maintained anymore. We also show time evolution of $C_{4}%
{(t)}$ for different values of $\tfrac{V}{\omega}$ and other parameters.

\section{Discussion}

In this section we discuss our results. We first look at the behavior of
$P_{(min)}^{4}$ as a function of $\tfrac{V}{\omega}$, see Fig.\ref{fig:4} We
observe that in weak field regime $P_{(min)}^{4}$ goes down as low as nearly
zero. However as the electric field strength is increased $P_{(min)}^{4}$
forms peaks centered at specific value of $\tfrac{V}{\omega}$ correponding to
close approach of two quasi-energies. This indicates the localization of
initial state at these points. We can understand this behavior terms of
behavior of quasi-energies. In Fig.\ref{fig:3} we present our result of
Floquet quasi-energy spectrum as a function of $\tfrac{V}{\omega}$.
\ Fig.\ref{fig:3(a)} is obtained by numerical diagonalization of one period
propagator $U[T,0]$ and in Fig.\ref{fig:3(b)} quasi-energies are obtained from
Eqn.(\ref{24}). The figure shows excellent agreement between perturbative
result with $(n=2)$ and the exact quasi-energies for strong and moderate
electric fields. For weak fields, the terms in ${H_{I}{(t)}}$ are not dominant
over $H_{t}$ terms and perturbation theory breaks down in weak field regime
which explains the decay of $P_{(min)}^{4}$ in this regime. The anti-crossings
in quasi energies suppress quantum coherent tunneling and we observe spikes in
$P_{(min)}^{4}$ occur at these anti-crossings. Hence it can be concluded that
the spikes in $P_{(min)}^{4}$ given in Fig.\ref{fig:3} correspond to
localization of the initial state which occur at the anti-crossing points of quasi-energies.

Since our system is initially in the triplet state $|{T_{0}}\rangle$, we
investigate the time evolution of this state through its amplitude $C_{4}%
{(t)}$ for different values of the ration $\tfrac{V}{\omega}$, Fig.\ref{fig:5}%
. One can clearly see that $C_{4}{(t)}$ remains unity for the entire time
evolution for curve(a) which is plotted at the exact anti-crossing point.
However, as we move away from the localization point, $C_{4}{(t)}$ decays
indicating that the initial state is no longer localized and tunneling is
possible, see Fig.\ref{fig:5} curves (b) and (c). Next we look at the effect
of increasing spin-orbit field strength $C_{4}{(t)}$. Time evolution of the
initial state is shown in Fig.\ref{fig:6} at the anti-crossing point
$\tfrac{V}{\omega}=14.79$ with spin-orbit strength set at $0.2$. The curve is
almost unity for entire evolution with a small oscillatory structure in
$C_{4}{(t)}$ between $1$ and $0.8$, indicating that this localization is
robust against spin-orbit field strength. Thus the periodic field forms a
well-established driven dynamics in the form of CDT even in the presence
spin-flip mechanism. For comparison we consider time evolution of $C_{4}{(t)}$
in the absence of AC field in Fig.\ref{fig:7}. We see that the state
oscillates with spin-flip nature transitions as the triplet state can only
couple to singlet states in our case. These oscillations are very rapid and do
not have a strictly periodic structure. However the inclusion of the AC field
prevents the spin from fliping and the state remains intact at as large values
as $\omega t=400,$ manifesting CDT, see Fig.\ref{fig:6}.

For completeness we also discuss the behavior of the initial state for
non-interacting electrons, given in Fig.\ref{fig:8}. In Fig. ${\ref{fig:8},}$
$\tfrac{V}{\omega}$ is set at different values but $U=0.$ The parameters are
chosen as ${(\varepsilon,W,V_{+},V_{-},n)=(0,0.1,0,0.05,0)}$ in units of
$\omega$. Red curve is the 4th zero of Bessel function i.e $11.79$. For black
and orange this ratio is $11$ and $12.5$. We find that at exact anti-crossing,
$C_{4}{(t)}$ remains unity even for non interacting electrons. The state shows
rapid oscillations at off resonance points as compared to interacting case
shown in Fig.\ref{fig:5}.

A remark on the strength of the spin-orbit field $\Omega$ is alse due here. We
refer to the values used in \cite{9}. For GaAs dots the spin-orbit coupling
$\Omega$, see Eqn.\ref{15}, using $W=10\mu eV$, an inter dot separation
$l=50nm$, and a spin-orbit length $\lambda_{SO}$ in the range $6-30\mu m$ we
get $|\Omega|=20-110neV$. However we have modeled spin-orbit field strength at
higher value than this typical value. This can be achieved by adjusting inter
dot separation $l=110nm$ \cite{25} and tunneling strength $W$.

To summarize, we have shown that the DQD system with two interacting electrons
under applied AC electric field can be used as a memory device. On the
localization points, the initial triplet state maintains its memory and
remains localized even though it is not the ground state of the system. Hence
at these points spin is conserved even in the presence of spin-orbit
interaction. Another interesting feature of this localization is that it is
robust against changes in parameters like spin-orbit field strength and
hopping amplitude in the system. It is expected that these results will be
important in future spintronic device applications based on spins in DQD.

\section{Acknowledgment}

K. Sabeeh would like to acknowledge the support of Higher Education Commission
(HEC) of Pakistan through project No. 20-1484/R\&D/09. K. Sabeeh would further
like to acknowledge the support of the Abdus Salam International Center for
Theoretical Physics (ICTP) in Trieste, Italy through the Associate Scheme.

\bigskip\clearpage%
%TCIMACRO{\FRAME{ftbpFU}{3.3797in}{2.3134in}{0pt}{\Qcb{The double quantum dot
%model and the coordinate system used: $S_{1}$ and $S_{2}$ denote the spin of
%the electron in the right and left quantum dots. The dots lie in $(\xi,z)$
%plane and tunneling is possible along $\xi$ direction. The dots are detuned by
%externally applied voltage and an AC field is applied to the dots as shown.}%
%}{}{1.eps}{\special{ language "Scientific Word";  type "GRAPHIC";
%width 3.3797in;  height 2.3134in;  depth 0pt;  original-width 3.333in;
%original-height 2.2736in;  cropleft "0";  croptop "1";  cropright "1";
%cropbottom "0";  filename '1.eps';file-properties "XNPEU";}} }%
%BeginExpansion
\begin{figure}
[ptb]
\begin{center}
\includegraphics[
height=2.3134in,
width=3.3797in
]%
{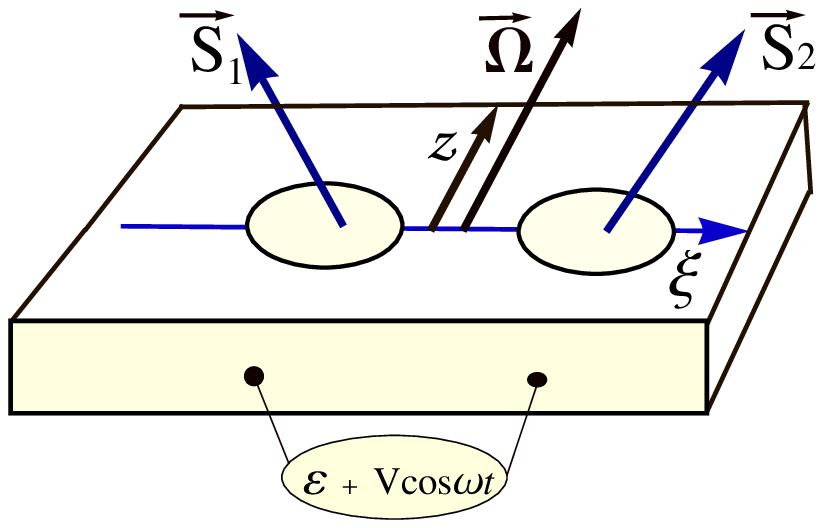}%
\caption{The double quantum dot model and the coordinate system used: $S_{1}$
and $S_{2}$ denote the spin of the electron in the right and left quantum
dots. The dots lie in $(\xi,z)$ plane and tunneling is possible along $\xi$
direction. The dots are detuned by externally applied voltage and an AC field
is applied to the dots as shown.}%
\end{center}
\end{figure}
%EndExpansion
\begin{figure}[ptb]
\centering
\subfigure[]{\label{fig:2(a)}\includegraphics[width=0.09\textwidth]{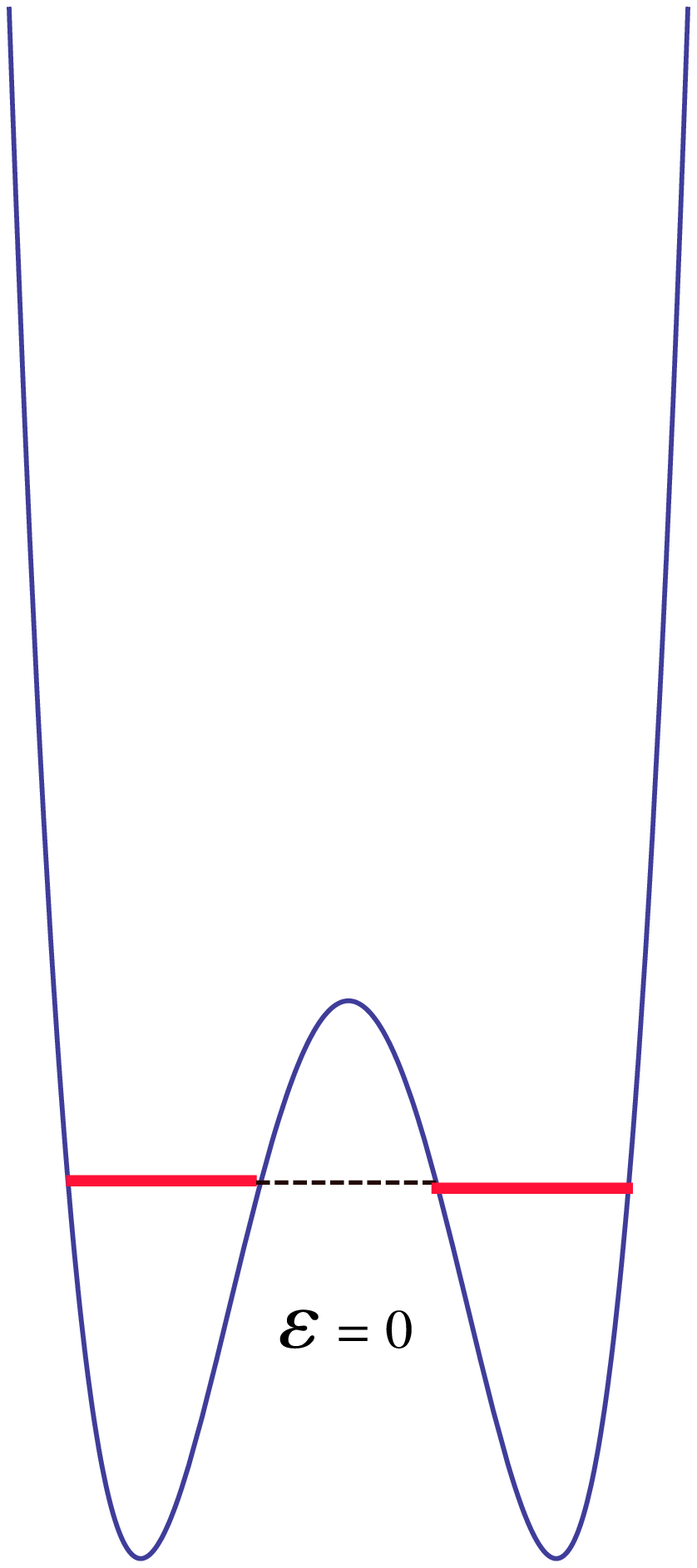}}
\subfigure[]{\label{fig:2(b)}\includegraphics[width=0.1\textwidth]{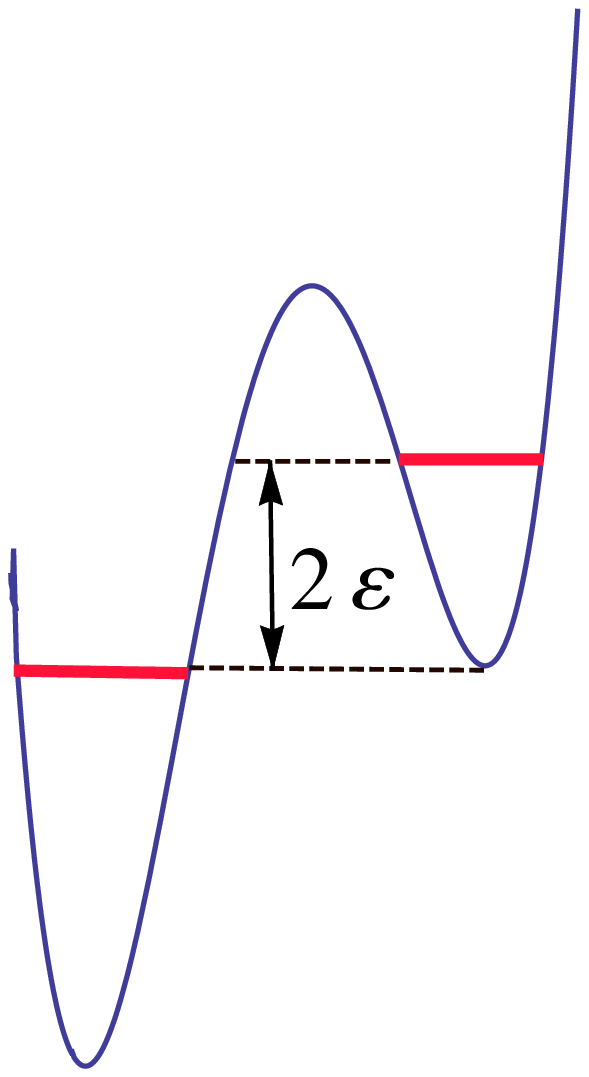}}\caption{$(a)$
At zero detuning an electron will have same energy in both the dots. $(b)$ For
nonzero detuning an electron on the left dot is at 2$\varepsilon$ lower energy
than an electron on the right dot.}%
\label{fig:2}%
\end{figure}\begin{figure}[ptb]
\centering
\subfigure[]{\label{fig:3(a)}\includegraphics[width=0.3\textwidth]{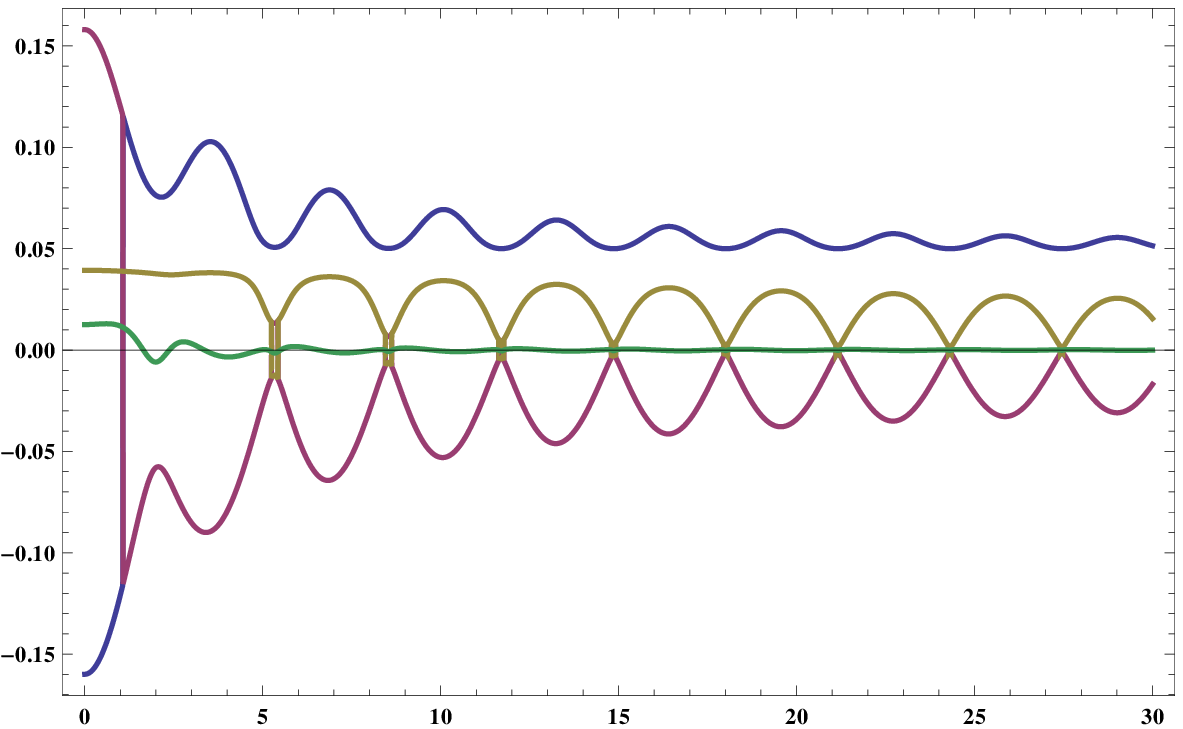}}
\subfigure[]{\label{fig:3(b)}\includegraphics[width=0.3\textwidth]{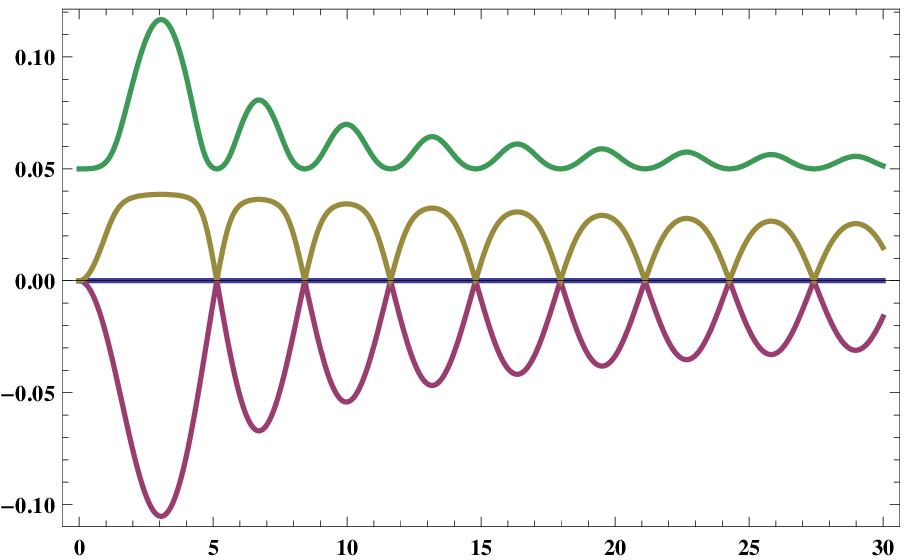}}\caption{Quasi-Energy
spectrum as a function of $\tfrac{V}{\omega}$:Figure ${\ref{fig:3(a)}}$ is the
perturbation theory result and ${\ref{fig:3(b)}}$ is the exact result. The
parameters are chosen as ${(U,\varepsilon,W,V_{+},V_{-},\Omega
)=(1,1,0.1,0,0.05,0.05)}$ in units of $\omega$. In the figure quasi-energies
and field strength $V$ are measured in units of $\omega$}%
\label{fig:3}%
\end{figure}\begin{figure}[ptb]
\centering
\includegraphics[width=0.6\linewidth]{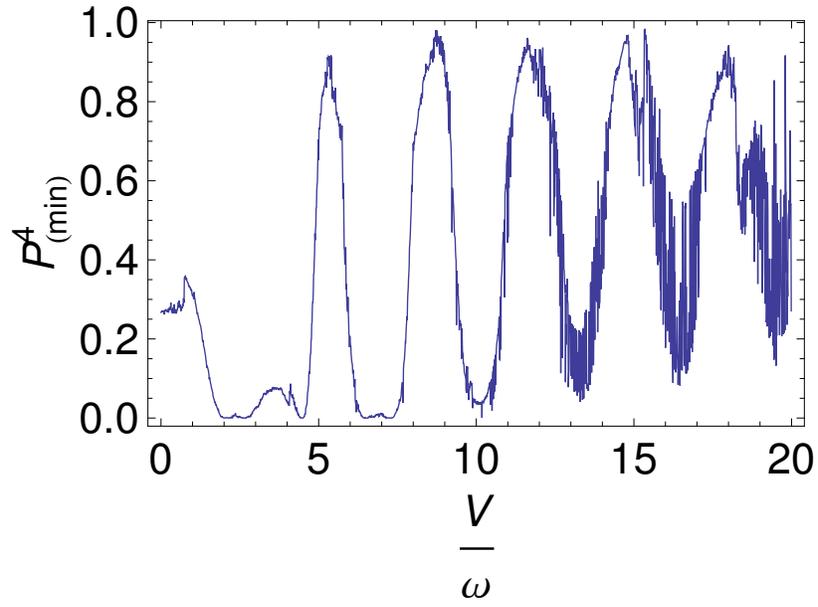}\caption{$P_{(min)}^{4}$ as a function of $\tfrac{V}{\omega}$ indicating localization behavior
of initial state}\label{fig:4}\end{figure}\begin{figure}[ptb]
\centering
\includegraphics[width=0.6\linewidth]{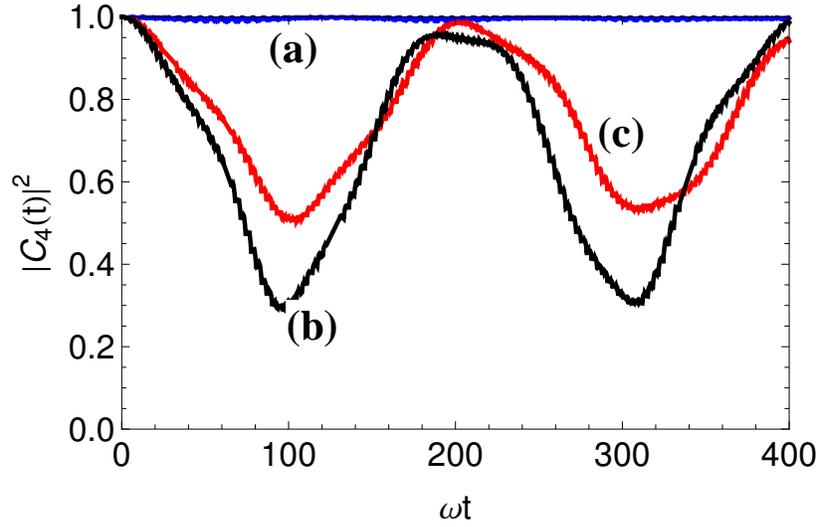}\caption{$C_{4}{(t)}$ as a
function of ${\omega t}$: (a) $\tfrac{V}{\omega}=14.79$ the exact
anti-crossing point. In this case $C_{4}{(t)}$ remains unity for all times.
Values of V are $V=13.5,16$, for (b),(c) respectively}%
\label{fig:5}%
\end{figure}\begin{figure}[ptb]
\centering
\includegraphics[width=0.6\linewidth]{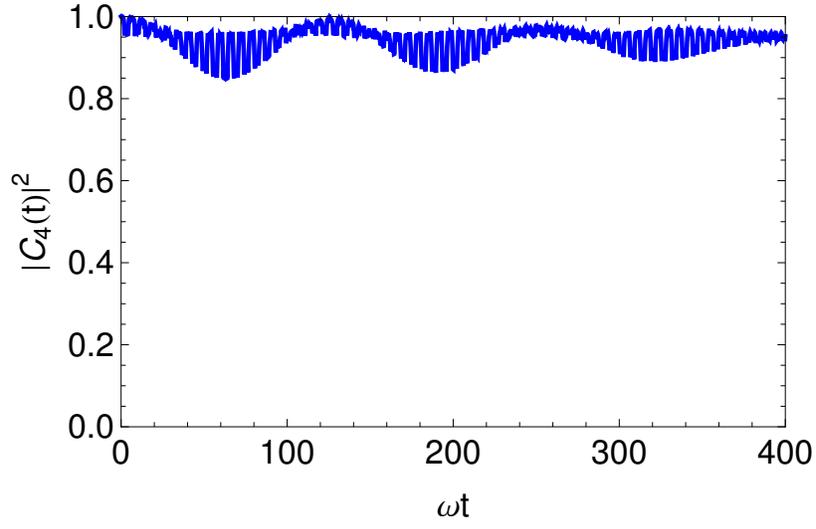}\caption{ The time evolution of
the initial state with $\Omega=0.2$ at the anti-crossing point $14.79$. Other
parameters are ${(U,\varepsilon,W,V_{+},V_{-})=(1,1,0.1,0,0.05)}$.}%
\label{fig:6}%
\end{figure}\begin{figure}[ptb]
\centering
\includegraphics[width=0.6\linewidth]{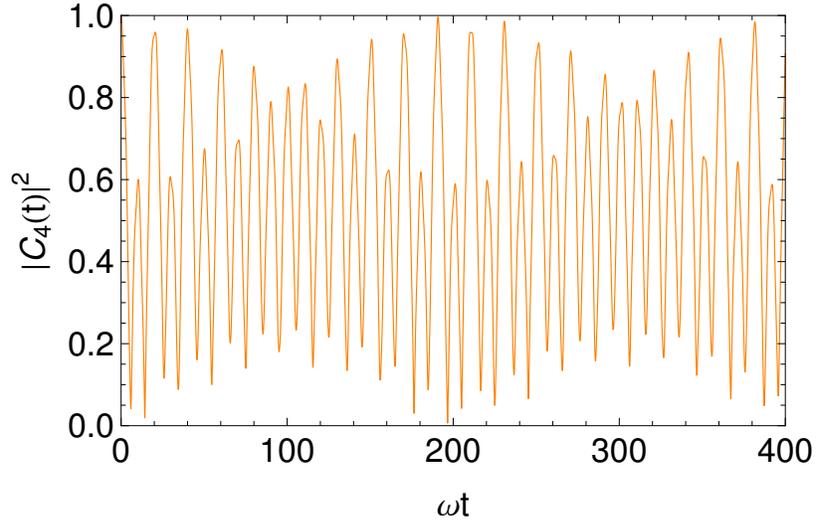}\caption{The time evolution of
$C_{4}{(t)}$ in the absence of AC field. Parameters are the same as in Fig.
\ref{fig:6}}%
\label{fig:7}%
\end{figure}\begin{figure}[ptb]
\centering
\includegraphics[width=0.6\linewidth]{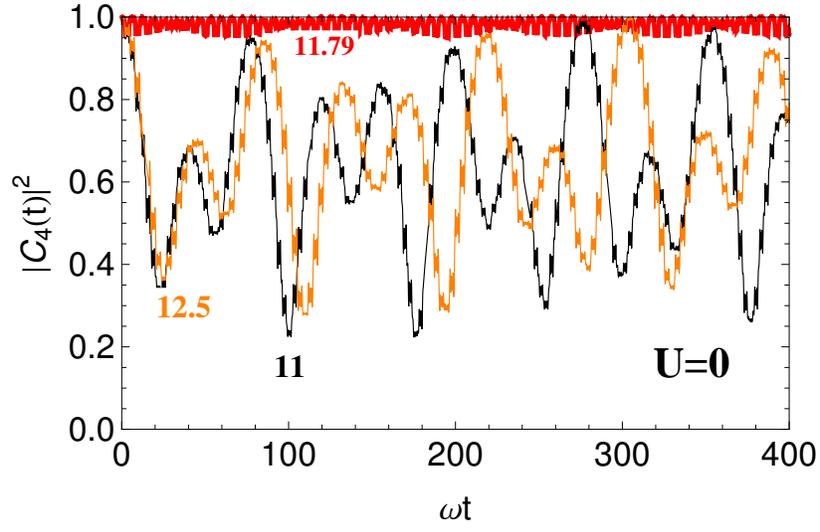}\caption{Behavior of initial
state for non interacting electrons for different values of $\tfrac{V}{\omega
}$ marked near the plots. Parameter choosen are ${(U,\varepsilon,W,V_{+}%
,V_{-},\Omega)=(0,0,0.1,0,0.05,0.2)}$. At exact anti-crossing $C_{4}{(t)}$
remains unity even for non interacting case}%
\label{fig:8}%
\end{figure}

\end{document}